%% file: ms.tex
\documentclass[12pt,preprint]{aastex}

\newcommand{\hi}{H\,{\sc i}}
\newcommand{\hii}{H\,{\sc ii}}
\newcommand{\ha}{H\,$\alpha$}
\newcommand{\kms}{km~s$^{-1}$}
\newcommand{\msol}{$M_{\odot}$}
\newcommand{\eg}{e.\,g.,}
\newcommand{\ie}{i.\,e.,}

\shorttitle{H{\sc i} in CenA dwarfs}
\shortauthors{Bouchard et al.}

\begin{document}

\title{Interstellar medium disruption in the Centaurus A group}

%\author{Antoine Bouchard}
\author{Antoine Bouchard\footnote{Current address: Universit\'e de Lyon 1, Centre de Recherche Astronomique de Lyon, Observatoire de Lyon, 9 avenue Charles Andr\'e, F-69230 Saint-Genis Laval, France; CNRS, UMR 5574 ; \'Ecole Normale Sup\'erieure de Lyon, Lyon, France.}}
\affil{Research School of Astronomy \& Astrophysics, Mount Stromlo Observatory, Cotter Road, Weston Creek, ACT 2611 Australia
\\{\rm and}\\
Australia Telescope National Facility, PO Box 76, Epping, NSW 1710, Australia}
%\affil{Universit\'e de Lyon 1, Centre de Recherche Astronomique de Lyon, Observatoire
%de Lyon, 9 avenue Charles Andr\'e, F-69230 Saint-Genis Laval, France ; CNRS,
%UMR 5574 ; \'Ecole Normale Sup\'erieure de Lyon, Lyon, France.}
\email{bouchard@obs.univ-lyon1.fr}

\author{Helmut Jerjen}
\affil{Research School of Astronomy \& Astrophysics, Mount Stromlo Observatory, Cotter Road, Weston Creek, ACT 2611 Australia}

\author{Gary S. Da Costa}
\affil{Research School of Astronomy \& Astrophysics, Mount Stromlo Observatory, Cotter Road, Weston Creek, ACT 2611 Australia}

\author{J\"urgen Ott\footnote{Bolton Fellow}}
\affil{Australia Telescope National Facility, PO Box 76, Epping, NSW 1710, Australia}

\begin{abstract}
We present the results of a 21\,cm neutral hydrogen (\hi) line detection
experiment in the direction of 18 low luminosity dwarf galaxies of the
Centaurus A group, using the Australia Telescope National Facility 64m
Parkes Radio Telescope and the Australia Telescope Compact Array.
Five dwarfs have \hi{} masses between $M_{\rm HI}=4\times10^5$ to
$M_{\rm HI}=2.1\times10^7$ \msol{} and $0.04<M_{\rm HI}/L_{\rm B}<1.81$
$M_{\odot} L_{\odot, {\rm B}}^{-1}$. The other 13 have upper-limits between
$M_{\rm HI}<5\times10^5$ and $M_{\rm HI}<4\times10^6$ \msol{} ($M_{\rm
HI}/L_{\rm B}<0.24$ $M_{\odot} L_{\odot, {\rm B}}^{-1}$). 
Two of the mixed-morphology dwarfs remain undetected in \hi{},
a situation that is in contrast to that of similar Local Group and Sculptor
group objects where all contain significant amounts of neutral gas.
There is a discontinuity in the \hi{} properties of Centaurus A group low
luminosity  dwarfs that is unobserved amongst Sculptor group dwarfs.  All
objects fainter than $M_{\rm B}=-13$ have either $M_{\rm HI}>10^7$\,\msol{}
or $M_{\rm HI}<10^6$\,\msol{}.  This gap may be explained by the ram
pressure stripping mechanism at work in this dense environment where all
galaxies with $M_{\rm HI}<10^7$\,\msol{} have been stripped of their gas.
The required intergalactic medium density to achieve this is
$\sim10^{-3}$\,cm$^{-3}$.

\end{abstract}


\section{Introduction}

Galaxy morphology is a static description of an evolving stellar system. It is
the visual manifestation of the physical processes that shaped the global optical light
distribution over a Hubble time.
It is the present day star formation, generating discreet pockets of young
bright stars and \hii{} regions, that leads to the irregular B-band
appearance of late-type dwarfs. Star formation activity, however,
can only operate until the resources for producing new stars in the form of
neutral hydrogen reservoirs, is exhausted.  This eventually results in an
object with a smooth, featureless light distribution (an early-type dwarf),
devoid of any significant amount of interstellar medium (ISM).  Whether or not these evolved
late-type galaxies will have the same properties as the present day
early-type dwarf galaxies is a much discussed matter \citep{mayer2001,
pedraz2002, simien2002, grebel2003, derijcke2003, derijcke2004, vanzee2004,
read2005}.  The evolutionary phase between the two major morphological
types provides observable objects
(known as transition-type dwarfs, dE/dIrr) with a continuum in both gas and stellar
properties between the two extremes: gas rich dwarf Irregulars (dIrr) and
gas poor dwarf Ellipticals (dE, also incorporating dwarf Spheroidals,
dSph and dwarf S0s, dS0). 

In essence, to be classified as ``early-type'', a dwarf galaxy only needs
to have had no recent star formation. This could, in principle be
independent of the presence or absence of \hi{} gas.  Indeed, while the
Local Group dEs exhibit a wide variety of star formation histories
\citep[\eg][]{grebel2001}, few show signs of recent star formation
\citep{tolstoy2004, babusiaux2005, olszewski2006} yet some are not
completely devoid of \hi{}.  The two early-type galaxies companions to M31,
NGC185 and NGC205 \citep{young1997} and the Sculptor dSph
\citep{carignan1998, bouchard2003} are examples \citep[see also
][]{blitz2000} where traces of \hi{} are detected ($M_{\rm
HI}~\sim~10^5$~\msol{}). Moreover, all known Local Group transition type
galaxies have been detected in \hi{}: LGS3 contains $M_{\rm HI}$ =
6$\times10^5$ \msol{} \citep{young1997, robishaw2002}, Phoenix has
1$\times10^5$ \msol{} \citep{st-germain1999, gallart2001}, Antlia has
7$\times10^5$ \msol \citep{barnes2001}, DDO210 has $2\times10^6$ \msol{}
and Pegasus has 5$\times10^6$ \msol{} \citep{lo1993}. The same observation
is made for
the four known transition type galaxies of the Sculptor group; these have
3$\times10^5$ \msol{} $<~M_{\rm HI}~<~10^6$ \msol{} \citep{bouchard2005}.
Two dS0 galaxies ESO384-G016, in the Centaurus A group, and NGC 59, in the
Sculptor group, have also been detected with $M_{\rm HI} > 10^6$ \msol
\citep{beaulieu2006}.  Similarly, in galaxy clusters, up to 15\% of
early-type dwarfs still have substantial amounts of \hi{} \citep[$M_{\rm
HI}~\gtrsim~10^7$ \msol{} ][]{conselice2003}.  It is therefore incorrect to
believe that only currently star forming or irregular galaxies contain
ISM.

The detection of \hi{} in early-type dwarf galaxies is an important step in
understanding any morphological evolution scenario. Gas depletion is still
viewed as the key factor driving the transition from late to early-type
\citep[\eg][]{conselice2003}.  Morphological properties of galaxies are
strongly correlated to their environment \citep{dressler1980, binggeli1987,
binggeli1990, vandenbergh1994b} and factors such as ram pressure
\citep{einasto1974} and tidal fields \citep{moore1996}, galactic winds
\citep{marcolini2004} and enhanced star formation efficiency
\citep{buyle2005} can, in principle, remove or exhaust the gas from a
late-type dwarf and force the transition. However, since ISM is detected in
some early-type dwarfs, there are other options one needs to study. (1)  Gas
depletion is not an end state, \ie{} gas can be accreted from various
sources such as stellar winds, intergalactic medium, etc.
\citep[\eg][]{mould1990}.
(2)  The gas is not depleted but is difficult to detect; \eg{} the ISM resides in an
ionised state \citep{mashchenko2004}. In these cases, dEs may still have
the means to form new stars and thus to oscillate between morphologies: some
could be quiescent mixed-morphology dwarfs.

By identifying and analysing low mass dwarf galaxies that contain \hi{}, we
can constrain this theory of evolution.  For the Local Group, the work has
been conducted by \citet{blitz2000} and \citet{bouchard2006a} who have
identified a number of early-types dwarfs possibly associated with \hi{}
emission.  In the Sculptor group, only one `genuine' early-type dwarf, \ie
one without any detected ISM, could be found \citep{bouchard2005}, in
accordance with morphological predictions \citep{jerjen2000a}.  This makes
the Sculptor group one of the rare environments where late-type dwarfs
vastly outnumber early-types: for example, down to $M_{\rm B}\,\sim-10$,
Sculptor has one early-type dwarf against 18 late-types
(\citealt{cote1997}, hereafter CFC97; \citealt{karachentseva1998},
hereafter KK98; \citealt{jerjen2000a}) while the Local Group has at least
17 early-type dwarfs against 16 late-type dwarfs \citep{mateo1998}.  The
Sculptor group has an elongated shape extending over $\sim3$ Mpc
\citep{jerjen1998, karachentsev2004} and is therefore more of a galaxy
`cloud' than a gravitationally bound entity; it does not have high galaxy
density regions. This makes environmental influences such as ram pressure
and tidal stripping largely ineffective and may explain the lack of
early-type objects.

The Centaurus A (CenA) group is a much denser environment than Sculptor.
CFC97 identified $\sim20$ dIrr galaxies in the CenA group using
photographic plates as well as \hi{} and \ha{} spectroscopy. In a
complementary study, 13 dwarf galaxy candidates of early and mixed
morphology were detected by \citet{jerjen2000b}, five of which had their
distances measured and membership confirmed by the surface brightness
fluctuation technique \citep{jerjen2000a}.  Several other catalogs and
studies describing galaxies of the CenA group have also been published
(KK98; \citealt{banks1999, karachentseva2000, karachentsev2002a,
karachentsev2006b}). In total, the CenA group contains at least 54 galaxies
and has NGC5128 (CenA) and NGC5236 (M83) at the center of its principal
density peaks. Of this number, $\sim50$ galaxies are dwarfs (with absolute $B$
magnitude fainter than $M_B > -18$): 32 are late-type, 14 are early-type
and 4 are of mixed-morphology. This rich group environment is, after
Sculptor, the next logical target to search for \hi{} rich dEs.

This paper presents the results of Parkes single-dish and Australia Telescope
Compact Array \hi{} observations in the direction of 18 CenA dwarf
galaxies.  The aim was to obtain a complete picture of \hi{} properties in
all known CenA group dwarf galaxies to a low \hi{} mass limit. The
following pages are divided as follow: Section 2 contains information on
the target selection and observations. The main results of this
investigation are found in Section 3 while the analysis and implications
are discussed in Section 4. Finally Section 5 presents the conclusions of
the paper.

\section{Observations}

\subsection{Sample selection}

\begin{figure}[tbp]
\begin{center}
\includegraphics[width=0.7\textwidth]{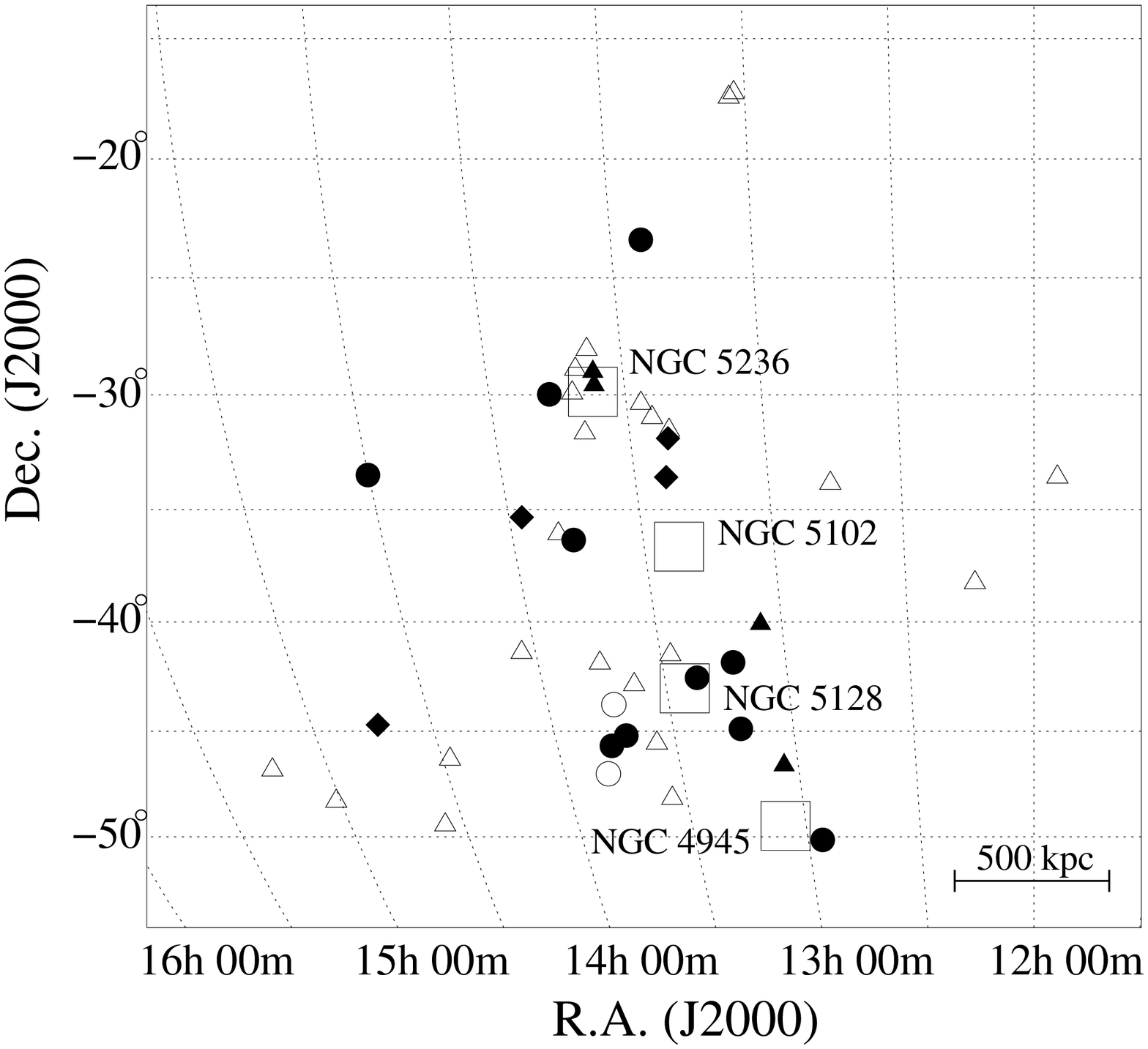}
\caption{Sky distribution of the CenA group galaxies. The large squares
mark the position of the major galaxies, while the circles represent early-type
dwarfs, triangles for late-type dwarfs and diamonds are used for
mixed-morphology dwarfs. Filled symbols represent the observed galaxies.}\label{cenagroup}
\end{center}
\end{figure}

The CenA galaxy group (Figure \ref{cenagroup}) was chosen because it
represents a
widely different environment to the low density Sculptor group, which we
studied earlier \citep{bouchard2005}.  The relatively large number of
galaxies, the presence of an active radio galaxy (CenA) and the
high overall density of the group makes it more akin to a cluster
environment.  The biggest advantage of the CenA group, however, is its
short distance to the Milky Way ($\sim4.1$ Mpc), which makes deep and detailed
\hi{} studies of its members possible.

Our aim is to study the \hi{} properties of galaxies that are the most
susceptible to mass-loss mechanisms. These mechanisms will operate most
effectively at the faint-end of the galaxy luminosity function.  From the
lists compiled by CFC97, \citet{jerjen2000b}, KK98, and
\citet{karachentsev2006b}, we see that the CenA group contains 35 known
dwarfs with $M_B \gtrsim -14$ (14 early-type, 17 late-type and 4 mixed
type) and, of these, we observed 18: 10 early-type, four late-type and all
four mixed-type.  The properties of these dwarfs, \ie{} the morphological
types, spatial positions, radial distances, apparent magnitudes, Galactic
extinctions, and optical heliocentric radial velocities,  are summarised in
Table~\ref{objects}.  While all these galaxies have previously been
observed with HIPASS \citep{barnes2001}, that all-sky survey lacked
sufficient resolution or sensitivity to constrain the properties
of these low mass objects. They were re-observed with either the
Parkes Radio Telescope or the Australia Telescope Compact Array. 

\input{tab1.tex}

\subsection{Parkes observations}

Using the same approach as \citet{bouchard2005},
the 64m Parkes Telescope\footnote{The Parkes telescope is part of the Australia
Telescope, which is funded by the Commonwealth of Australia for operation as a
National Facility managed by the CSIRO.} was employed to obtain high spectral
resolution \hi{} line spectra in the direction of the early-type dwarfs AM1343-452,
ESO269-G066, CenA-dE1 and CenA-dE4. The observations were conducted in
February 2005 (project P475). The Multibeam instrument in MX
(beam-switching) mode and the narrowband correlator with the MB7\_8\_2048
settings, provided a bandwidth of 8 MHz divided in 2048 channels and 2
polarisations. The central frequency was set at 1417 MHz resulting in an \hi{}
velocity coverage from -100 to 1500 \kms, with a channel width of 0.82
\kms. The beam size is 14.1$^{\prime}$ or $18$\,kpc at a radial
distance of 4.3\,Mpc. In
MX mode only the seven inner-most horns of the multibeam detectors are used.
At all times, one of the beams was kept on-source while
the six others observed adjacent sky to provide bandpass calibration, alternating
the on-source beam every two minutes. The total
integration times %can be found in Table \ref{hidetections}.
ranged from 350 to 650 minutes (see Table~\ref{hidetections}).

These data were reduced with the LiveData data reduction pipeline. The median
of the Tukey-smoothed bandpass was used for calibration and the data were
gridded with Gridzilla after averaging the two polarisations. LiveData and
Gridzilla are part of the AIPS++ software package. The \emph{mbspect} robust
polynomial fitting algorithm from the MIRIAD software package was used to fit
low order polynomials and subtract residual baseline ripples.

\subsection{ATCA observations}

The Australia Telescope Compact Array (ATCA)\footnote{The Australia
Telescope Compact Array is part of the Australia Telescope, which is funded
by the Commonwealth of Australia for operation as a National Facility
managed by CSIRO.} was used to obtain \hi{} line maps and
spectra in the direction of the 13 galaxies:
ESO269-G037, [CFC97]Cen8, AM1339-445 and ESO384-G016 with the 1.5A array
configuration, [CFC97]Cen6 and CenA-dE5 with the 750D configuration,
ESO219-G010, CenA-dE2, AM1320-230 and CenA-dE3 using the EW367
configuration and
SGC1319.1-4216, UGCA365 and ESO272-G025 with the EW352 configuration.
In addition, observations in the direction of [CFC97] Cen5 were also
conducted but are not presented here. This galaxy is a background spiral,
not associated with the CenA group \citep{bouchard2004}.  

The arrays were chosen so that the final synthesised beam would encompass a maximum of the anticipated flux structure while not being much larger. Precisely, it was the different distances to each object and their respective morphological appearances that were taken into account. This was done to avoid over-resolving the sources which would have led to detection difficulties whilst also avoiding the dilution of the \hi{} signal with the surrounding noise. The observations
were carried out in 2003 February and March and in 2004 March and April
(project C1133).

The FULL\_4\_1024-128
correlator configuration was employed with the central frequency of 1417 MHz for an
\hi{} line velocity coverage from 200 \kms{} to 1000 \kms{} or of 1416 MHz,
covering from 400 \kms{} to 1200 \kms{}. The spectra were divided in
1024 channels each of 0.82 \kms. For each observing session the radio
continuum source 1934-638 was observed for 10 minutes for use as a flux
and bandpass calibrator. A phase calibrator was also observed for typically
5 minutes every 40 minutes of on-source integration. Total integration
times ranged between 270 and 650 minutes (See Table~\ref{hidetections}).

The data were reduced with the MIRIAD software package using standard
procedures. The continuum was fitted and subtracted using a low order polynomial.
Each data cube was produced using ``natural'' weighting of
baselines, CLEANed and RESTORed to a Gaussian beam of the same size as the
main lobe of the
synthesized beam. 

\section{Results and \hi{} properties}

\input{tab2.tex}

Our program has detected \hi{} emission in five galaxies: the dIrr ESO269-G037, the
dIrr [CFC97]Cen6, the dIrr UGCA365, the dE/dIrr ESO384-G016 and the dE/dIrr
ESO272-G025. Line emission maps can be found in Figures \ref{atca} and
\ref{atca384}; the spectra in Figure \ref{spectra}.  Of the 11
galaxies that were not detected, two are of mixed morphology: CenA-dE2 and
[CFC97]Cen8 while one is a dIrr: [KK98]208. The rest are early-type
dwarfs.

In Table~\ref{hidetections} we compile the \hi{} detection parameters or
upper-limits on the \hi{} content.
For each galaxy, we list the telescope that was used and the
total on-source integration time, T$_{\rm int}$ in minutes.
For the detected galaxies, the heliocentric
radio-velocities $V_{\odot}$, and velocity dispersions $\sigma_V$ of the
gas are listed.
The total \hi{} flux, in Jy \kms{}, is given by:
\begin{equation}
S_{\rm HI}~=~\int S_v \,\, {\rm d}v
\end{equation}
\noindent where $S_v$ is the flux value within each channel of velocity
width ${\rm d}v$.
The \hi{} mass $M_{\rm HI}$, in solar units $M_{\odot}$, was calculated
using the standard formula:
\begin{equation}
M_{\rm HI}~=~2.356\times10^5~D^2~S_{\rm HI}
\end{equation}
\noindent where $D$ is the radial distance in Mpc. When $D$ was
unavailable from the literature, a value of $4.3\pm0.7$ Mpc was adopted, which corresponds to the
mean and standard deviation of the distances to all galaxies in the CenA group
\citep{karachentsev2006b}.

The \hi{} mass to blue luminosity ratio $M_{\rm HI}/L_{\rm B}$, in solar
units $M_{\odot} L_{\odot, {\rm B}}^{-1}$ is:
\begin{equation}
M_{\rm HI}/L_{\rm B}~=~ 1.5\times10^{-7} ~ 10^{\,0.4\,(m_{\rm B} - A_{\rm B})} ~ S_{\rm HI}
\end{equation}
\noindent in which $m_{\rm B}$ is the apparent integrated B-band magnitude
of the object and $A_{\rm B}$ is the B-band Galactic extinction value along
the line of sight (see Table~\ref{objects}). The absolute magnitude of the
sun was taken as $M_{\odot, {\rm B}}=5.5$ \citep{bessell1998}. 
The quoted errors are the results of quadratic error propagation and, when no error were mentioned for $m_{\rm B}$, these values were taken as $\pm0.1$\,mag.

Where the galaxies were not detected, Table~\ref{hidetections} lists
upper-limits for $S_{\rm HI}$, $M_{\rm HI}$, and $M_{\rm HI}/L_{\rm B}$.
These upper-limits were calculated with the same above equations but using
a value of $S_v$ equal to three times the RMS in the spectra integrated
over 10 \kms{}. The RMS was typically $\sim4$\,mJy\,beam$^{-1}$ in ATCA
observations and $\sim1$\,mJy\,beam$^{-1}$ in Parkes data.

We should note, however, that while ESO269-G066 is listed as not detected
in Table~\ref{hidetections}, it contains a prominent \hi{} feature in its
spectrum % Figure \ref{spectra_eso269} shows a significant detection 
at $v_{\odot}$~=~231$\pm$1 kms$^{-1}$ with $S_{\rm HI}\sim0.33$ Jy \kms{}
(Figure \ref{spectra_eso269}).  This feature was not reported by
\citet{beaulieu2006} who observed the same object with the Green Bank
Telescope, covering a heliocentric velocity range from 600 to 1000 \kms{};
they have placed an upper-limit of $M_{\rm HI}\,<\,1.6\times10^6$\,\msol{}
(corrected to a distance of 4.05\,Mpc).  The optical velocity of this
object was established to be $v_{\odot}$~=~784$\pm$31 \kms, measured from
Balmer absorption lines \citep{jerjen2000a}. It therefore seems unlikely that
this \hi{} feature is associated with the dwarf. A systemic velocity of
231\,\kms{} would also be inconsistent with a CenA group association, most
other objects have velocities greater then 500\,\kms. Inspection of the High
Velocity Cloud catalog \citep{putman2002} reveals many HVCs in the
vicinity of ESO269-G066 at $v_{\odot}\sim200$ \kms{} and one may have been
caught in the 14\arcmin{} beam of the Parkes telescope. We conclude that no
\hi{} is associated with this galaxy, which therefore has
$M_{\rm HI}~<~10^5$ \msol.

\section{Analysis}

\subsection{Early-type dwarfs}
The first interesting result from our \hi{} study that should be pointed
out is that none of the ten observed dwarfs classified as dE have been
detected in \hi{}.  The $M_{\rm HI}$ upper-limits range between
$1.0\,-\,6.5\times10^5$\msol{} and the $M_{\rm HI}/L_{\rm B}$ upper-limits
are between $0.002\,-\,0.20\,M_{\odot} L_{\odot, {\rm B}}^{-1}$.  To
understand the implications of the $M_{\rm HI}/L_{\rm B}$ results, one
should consider two extreme examples: CenA-dE5 and ESO269-G066.  While both
have roughly similar \hi{} mass upper-limits, $M_{\rm
HI}\,=\,3.4\times10^5$\,\msol{} and $M_{\rm
HI}\,=\,1.0\times10^5$\,\msol{}, respectively, they have very different
$M_{\rm HI}/L_{\rm B}$ ratios. 

CenA-dE5 is the faintest galaxy of our sample and has an absolute magnitude
$M_{\rm B}$ = -10.1 (for an assumed distance of 4.3 Mpc).  Although this
object is brighter than some Local Group dwarfs (\eg{} Ursa Minor, $M_{\rm
B}$ = -7.6 \citealt{mateo1998}), it is amongst the faintest known galaxies
in the local universe. It has $M_{\rm HI}/L_{\rm B}~<~0.2$ $M_{\odot}
L_{\odot, {\rm B}}^{-1}$.
At the bright end of our optical luminosity distribution, ESO269-G066 has
$M_{\rm B}$ = -13.8 and $M_{\rm HI}/L_{\rm B}~<~2\times10^{-3}$ $M_{\odot}
L_{\odot, {\rm B}}^{-1}$.

As a first approximation, both these galaxies can be considered as being
dominated by an old and metal poor stellar population \citep[see][]{jerjen2000a}.  The mass loss expected from the evolution of such a
population is of the order of $0.1\,M_{\odot} L_{\odot, {\rm B}}^{-1}$ over
a Hubble time \citep{bouchard2005}.  It becomes immediately obvious that
while the $M_{\rm HI}/L_{\rm B}$ upper-limit for CenA-dE5 does not exclude
such mass loss material being in the form of \hi{}, it is very stringent on the state of
ESO269-G066.  In this latter case, any neutral gas build-up is, at best,
insignificant. The upper-limit on $M_{\rm HI}/L_{\rm B}$ actually requires
that the ISM in this galaxy is either completely ionised or, more likely,
has been regularly and thoroughly swept out of the dwarf.  For comparison
purposes, the Local Group dE galaxies NGC185 and NGC205, both satellites of
M31 have $M_{\rm HI}/L_{\rm B}\,=\,4\times10^{-3}$ and
$3\times10^{-3}\,M_{\odot} L_{\odot, {\rm B}}^{-1}$, respectively
\citep{young1997}.

\subsection{Mixed-type dwarfs}

Of the four mixed-type dwarfs that were observed, two were detected in
\hi{}: the dE/dIrr ESO384-G016 with $M_{\rm
HI}~=~(4.4\pm0.8)\times10^6$ \msol{},  $M_{\rm HI}/L_{\rm
B}~=~0.13\pm0.02$ $M_{\odot} L_{\odot, {\rm B}}^{-1}$ and the dE/dIrr
ESO272-G025 with $M_{\rm HI}~=~(7\pm2)\times10^6$ \msol{} and $M_{\rm
HI}/L_{\rm B}~=~0.10\pm0.01$ $M_{\odot} L_{\odot, {\rm B}}^{-1}$.
The two others, the dE/dIrr CenA-dE2 and the dIrr/dE [CFC97]Cen8 have
upper-limits of $M_{\rm HI}\,<\,5.1\times10^5$\,\msol{}, $M_{\rm HI}/L_{\rm
B}\,<\,0.24$\,$M_{\odot} L_{\odot, {\rm B}}^{-1}$ and $M_{\rm
HI}\,<\,5.5\times10^5$\,\msol{}, $M_{\rm HI}/L_{\rm
B}\,<\,0.17$\,$M_{\odot} L_{\odot, {\rm B}}^{-1}$, respectively.

We note that \citet{beaulieu2006} previously observed ESO384-G016 with the
ATCA and found $M_{\rm HI}~=~(5.6\pm0.3)\times10^6$ \msol{}, in good
agreement with our result. They had a longer integration time 
which resulted in a higher signal-to-noise ratio and, most probably, a more accurate result.
%and used the 1.5C 
%array configuration (we used 1.5A) which probably made their observations
%slightly more sensitive to the
%different structures of the source. 
These authors have also found that
single dish observation yielded $M_{\rm HI}~=~(6.5\pm0.1)\times10^6$
\msol{}, which highlights the fact that approximatively 25\% of the flux
resides in scales inaccessible to our array configuration. We also note
that ESO272-G025 had previously been detected in \ha{} but not in \hi{}
\citep{cote1997}.

At first sight, this situation may seem contrasting to that of both the
Local Group and the Sculptor group where all mixed-morphology galaxies have
been detected in \hi{}.  On closer inspection, however, Local Group
mixed-morphology dwarfs have $M_{\rm HI}/L_{\rm B}$ between 0.21 and 1.4
$M_{\odot} L_{\odot, {\rm B}}^{-1}$ \citep{st-germain1999, blitz2000, bouchard2006a}, while in the Sculptor group these
objects have $M_{\rm HI}/L_{\rm B}$ between 0.08 and 0.18 $M_{\odot}
L_{\odot, {\rm B}}^{-1}$ \citep{bouchard2005}. The  $M_{\rm HI}/L_{\rm B}\,<\,0.24\,M_{\odot}
L_{\odot, {\rm B}}^{-1}$ constraint imposed on the two non-detected CenA
dwarfs does not exclude \hi{} contents similar to those of Sculptor mixed morphology
dwarfs.

\subsection{Late-type dwarfs}

It comes with some surprise that not all observed late type dwarfs were
detected in \hi{}. 
While ESO269-G037,
[CFC97]Cen6 and UGCA365 all have considerable amounts of ISM, 
$M_{\rm HI}~=~(4\pm1)\times10^5$ \msol{} and
$M_{\rm HI}/L_{\rm B}~=~0.04\pm0.02$ $M_{\odot} L_{\odot, {\rm B}}^{-1}$,
$M_{\rm HI}~=~(42\pm7)\times10^6$ \msol{} and
$M_{\rm HI}/L_{\rm B}~=~1.81\pm0.08$ $M_{\odot} L_{\odot, {\rm B}}^{-1}$
and
$M_{\rm HI}~=~(18\pm3)\times10^6$ \msol{} and
$M_{\rm HI}/L_{\rm B}~=~0.56\pm0.02$ $M_{\odot} L_{\odot, {\rm B}}^{-1}$,
respectively, [KK98]208 was not detected with limits of
$M_{\rm HI}~<~4.1\times10^6$ \msol{} and
$M_{\rm HI}/L_{\rm B}~<~0.05$ $M_{\odot} L_{\odot, {\rm B}}^{-1}$.
From empirical evidence, we would have expected most dIrr galaxies to have 
$M_{\rm HI}/L_{\rm B}$ between 0.1 and 10 $M_{\odot} L_{\odot, {\rm B}}^{-1}$
at
$M_{\rm B}\,=\,-14$ \citep{warren2006}.

The galaxy [KK98]208 was added to the sample after the observations as it resides
only 30\arcmin{} away from UGCA365. %However, although radio interferometry
%techniques allow the field of view to be extended to almost any size, the
%sensitivity of the off-axis observation is far from optimal. 
The upper-limit for $S_{\rm HI}$ and correspondingly for $M_{\rm HI}$ is higher
than for any other galaxy of our sample (see Table~\ref{hidetections})
and, with $M_{\rm B}~=~-14.2$, this galaxy is optically brighter than most
other objects in our sample. This makes the $M_{\rm HI}/L_{\rm B}$ ratio
limit of 0.05 $M_{\odot} L_{\odot, {\rm B}}^{-1}$ surprisingly %well constrained.
low for a dIrr. No dwarf galaxy in the \citet{koribalski2004} sample has values
this low \citep[also see][]{warren2006}.

[KK98]208 also happens to share a similar line of sight to the spiral
NGC5236 (M83), which has \hi{} extending up to and beyond the spatial
position of [KK98]208. \citet{huchtmeier2000} claim to have detected the
dwarf at $V_{\odot}$ = 400 km s$^{-1}$, but this detection is confused with
the 21\,cm signal from NGC5236. The northern spiral arm of NGC5236 is
easily detected in our data at the velocity mentioned by
\citet{huchtmeier2000}, but there are no signs of kinematically-decoupled
\hi{} emission near the dwarf. It is also worth noting that
\citet{karachentsev2002a} found, in the colour-magnitude diagram of this
dwarf, a predominantly old stellar population
typical of a dE, %although the galaxy is of irregular
%morphology \citep{karachentseva1998}. 
in disagreement with its irregular
morphology (KK98). However, its extreme low surface brightness makes
classification on morphological grounds intrinsically difficult \citep[see
Figure 3 from][online material]{karachentsev2002a}. Both the stellar
population and the \hi{} properties of this object clearly favour an early-type
classification.

It is also interesting to note that the dIrr galaxy ESO269-G037 has $M_{\rm
HI}/L_{\rm B} = 0.04$\,$M_{\odot}\,L_{\odot, {\rm B}}^{-1}$. This value is
low for a dwarf of late-morphology. Based on stellar photometry,
\citet{karachentsev2002a} argued that this galaxy should actually be considered as a
dSph. The object may also harbour a small population of blue stars, which
would be consistent with the \hi{} result.

\subsection{\hi{} displacement and ram pressure}

\begin{figure}[tbp]
\begin{center}
\includegraphics[width=0.95\textwidth]{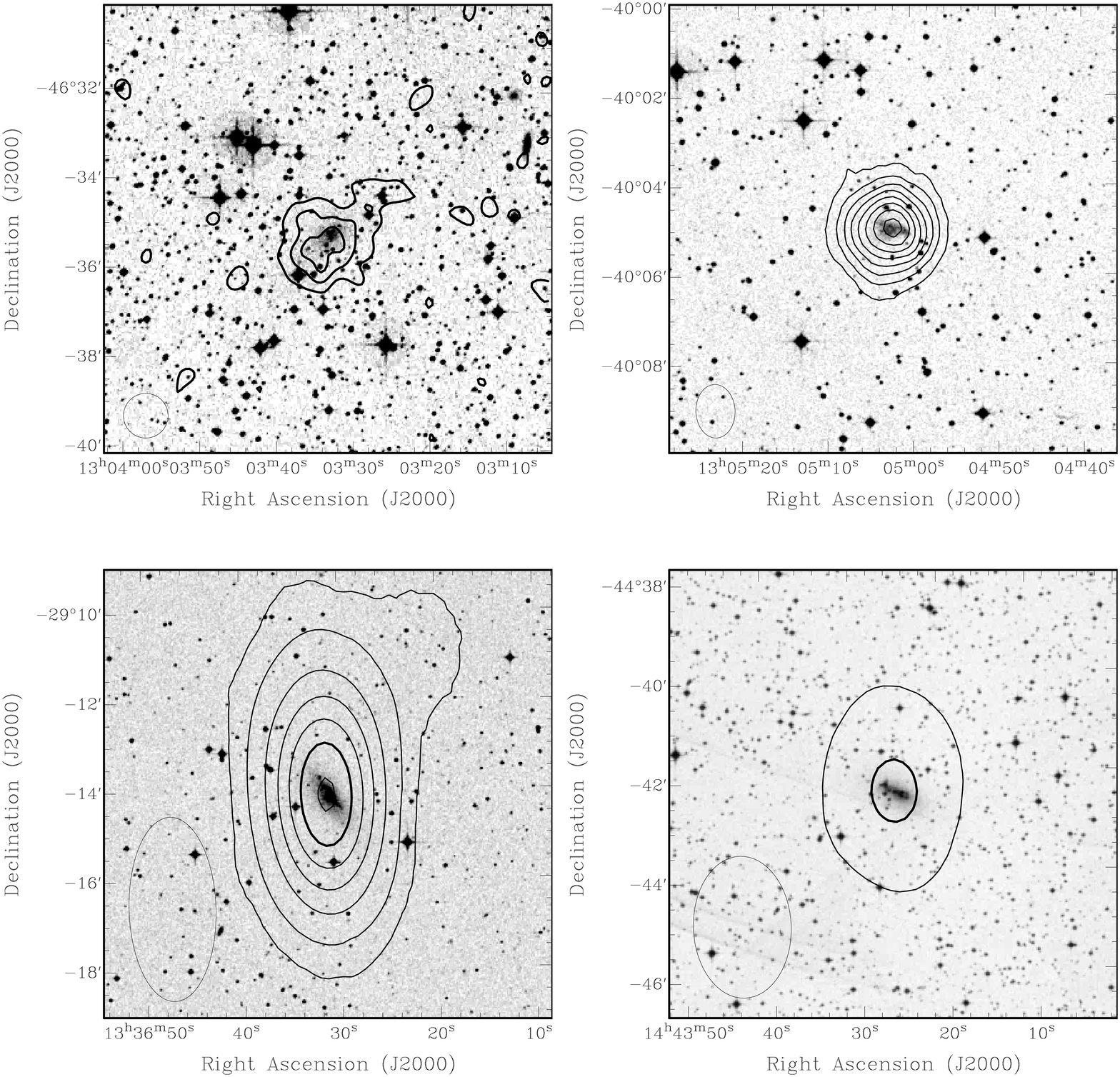}
\caption{\hi{} distribution map (\emph{contours}) overlayed on
10$^{\prime}\times10^{\prime}$ DSS images (or $\sim12$\,kpc$\times12$\,kpc
at an average distance of 4.1\,Mpc) of the dIrr ESO269-G037
(top-left), the dIrr {[CFC97]Cen06} (top-right), the dIrr UGCA365
(bottom-left) and dE/dIrr ESO272-G025 (bottom-right). The contour levels
are from $5\times10^{19}$ cm$^{-2}$ to $5\times10^{20}$ cm$^{-2}$ by steps
of $5\times10^{19}$ cm$^{-2}$ except for ESO269-G037 where contours are of
$2.5$, $5.0$ and $7.5\times10^{19}$ cm$^{-2}$. The ellipses in the
bottom-left corner of each frame is the synthesised beam size:
60\arcsec$\times$60\arcsec{} for ESO269-G037, 72\arcsec$\times$53\arcsec{}
for {[CFC97]Cen6}, 247\arcsec$\times$117\arcsec{} for UGCA365 and
171\arcsec$\times$118\arcsec{} for ESO272-G025.  } \label{atca}
\end{center}
\end{figure}

\begin{figure}[tbp]
\begin{center}
\includegraphics[width=0.45\textwidth]{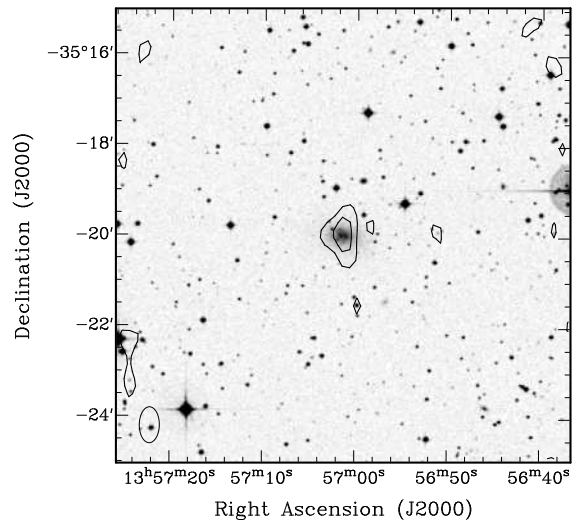}
\caption{\hi{} distribution map (\emph{contours})
overlayed on a 10$^{\prime}\times10^{\prime}$ DSS image (or  $\sim12$\,kpc$\times12$\,kpc
at an average distance of 4.1\,Mpc) of the dE/dIrr
ESO384-G016. The contour levels are of $1\times10^{19}$ and
$2\times10^{19}$ cm$^{-2}$.  The ellipse in the bottom-left corner is the
synthesised beam size: 47\arcsec$\times$26\arcsec{}. \hi{} was previously
detected in this galaxy by \citet{beaulieu2006}.}\label{atca384}
\end{center}
\end{figure}

\begin{figure}[tbp]
\begin{center}
\includegraphics[width=0.7\textwidth]{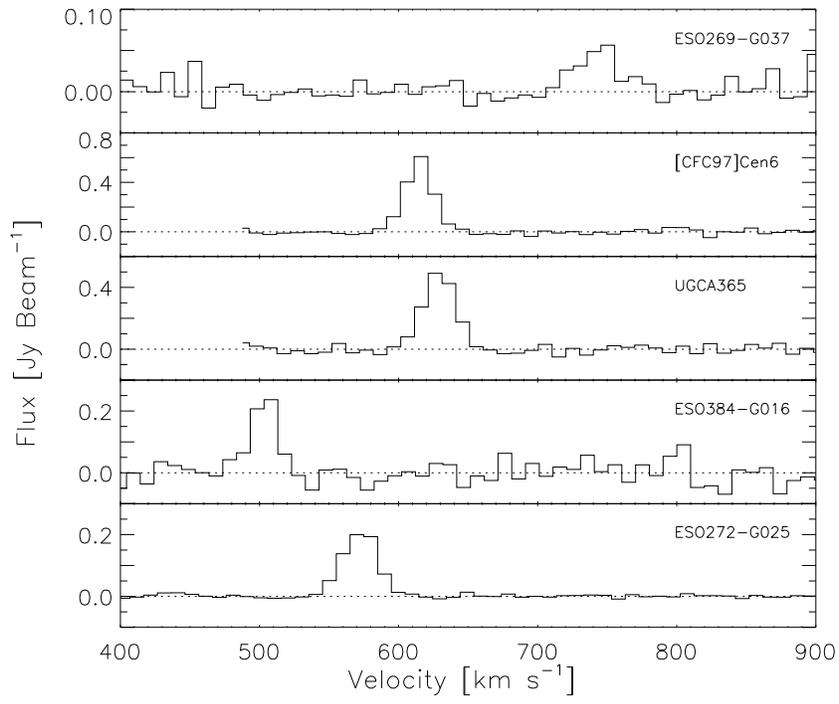}
\caption{ATCA \hi{} spectra for ESO269-G037, {[CFC97]Cen06},
UGCA365, ESO384-G016 and ESO272-G025. 
The spectra were binned to a 10 km\,s$^{-1}$ velocity resolution.}\label{spectra}
\end{center}
\end{figure}

\begin{figure}[tbp]
\begin{center}
\includegraphics[width=0.70\textwidth]{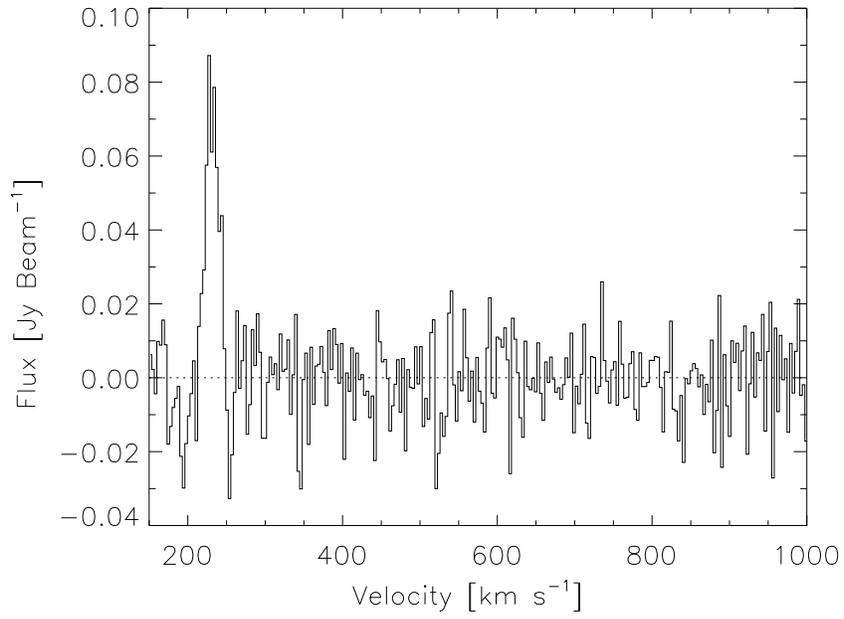}
\caption{Parkes \hi{} spectra for ESO269-G066. 
The spectra was binned to a 3.3 km s$^{-1}$ velocity resolution.}\label{spectra_eso269}
\end{center}
\end{figure}

Angular displacement of the \hi{} with respect to the optical center is
observed in a number of dwarf galaxies. In the Local Group, this is seen in
two of the five mixed-morphology galaxies, Phoenix and LGS 3
\citep{st-germain1999, gallart2001, robishaw2002}, while in the Sculptor
group, this is likely the case for two of the three investigated
mixed-morphology dwarfs.  In the CenA group, we only detect a possible
\hi{} displacement in ESO269-G037, but the low signal to noise
of the map presented in Figure \ref{atca} makes it difficult to accurately
determine if the gas is really offset with respect to the optical centre.

We do, however, detect a discrepancy between the \hi{} velocity
$v_{\odot}~=~502.9\pm2.4$ \kms{} (consistent with the measurement of
\citealt{beaulieu2006}) and the optical velocity $v_{\odot}~=~561\pm32$
\kms{} \citep{jerjen2000a} for ESO384-G016. While \citet{beaulieu2006}
argued that the \hi{} velocity is probably more accurate than the optical,
they also suggested that, based on the \hi{} distribution, this galaxy may
be experiencing ram pressure while falling into the group.
\citet{jerjen2000a} noted that this galaxy has an old and
metal-poor stellar population and the optical spectra shows no sign of
current star formation. Ram pressure stripping may be able to displace the
gas while keeping it with a smooth distribution \citep{gallart2001}, not
necessarily triggering star formation. 
% ***** Helmut's bit *****
Moreover, the \hi{} is at a lower velocity than the optical, a situation we
would expect if ESO384-G016, at $D = 4.2$\,Mpc, is falling onto NGC5236 
at $D = 5.1$\,Mpc \citep{karachentsev2006b}.
%One would expect the 60 \kms{} velocity
Since the mechanism causing this offset is unlikely to be
aligned with the radial direction, the velocity difference should also
%difference to 
have produced a measurable angular offset if it is the
result of a gentle `push' or have produced star forming regions if
the event was more violent. Alternatively, it may be that the gas is being
compressed and star formation is about to start. This object would then
evolve back toward a late-type morphology.  The spectral information
available for this object is of too low signal-to-noise. Further
observations will be required to investigate whether the \hi{} line shape
may have been influenced by ram pressure.

The Eastern \hi{} extension detected by \citet{beaulieu2006} in ESO384-G016
is similar to the north-western \hi{} extension of UGCA365 (Figure
\ref{atca}). This latter galaxy is situated %a projected angular distance of $\sim0.9^{\circ}$ or 
$\sim81$\,kpc three dimensional distance from NGC5236
(M83), a separation reminiscent of that of the Large and Small Magellanic
Clouds to the Milky Way.  It is possible that this object is
experiencing ram pressure and tidal stripping as it travels on its orbit
around M83.

\subsection{CenA environment and dwarf evolution}

The CenA group is a relatively high density environment providing external
conditions which can influence the evolution of its group members.  In
fact, one of the most important features of this environment is the
presence of the active galaxy NGC5128 (CenA).  The radio-lobes of NGC5128
may be dramatically affecting nearby objects (see Figure \ref{pos_cena}).
These regions of hot ionised plasma measure 9\degr{} ($\sim600$ kpc) in the
north-south direction and 3\degr{} ($\sim200$ kpc) east-west
\citep[\eg][]{junkes1993}. This may have affected the evolution of nearby
dwarfs. 

There are six galaxies projected near NGC5128.  These include the three
early type dwarfs, CenA-dE1, SGC1319.1-4216 and ESO269-G066, none of which
were detected in \hi{} (Table~\ref{hidetections}).  The three other objects
are gas rich (CFC97, \citealt{koribalski2004}).  The spiral ESO270-G017
(also known as Fourcade-Figueroa) and the irregular NGC5237 are believed to
be the remnants of a close interaction between a spiral galaxy and NGC5128,
which may have been at the origin of the latter's observed dust lane
\citep{dottori1973, thomson1992}. Finally, the dIrr ESO324-G024 is known
to have \ha{} and H$\beta$ emission, both signs of active star formation
\citep{lee2003}.

In the case of ESO269-G066, their is a good agreement in radial distance
with NGC5128: $D~=~3.84\pm0.35$ Mpc, \citep{rejkuba2004} compared to
$4.05\pm0.53$ Mpc for ESO269-G066 \citep{jerjen2000a}. The relative line
of sight velocity is $\Delta v\sim240$ \kms{} \citep[$v_{\odot}({\rm
NGC5128})~=~543\pm2$ \kms, ][]{israel1998}. It is possible that the orbit
of the dwarf brings it regularly through the lobes of NGC5128. The very low
limit on the \hi{} mass to light ratio of ESO269-G066, $M_{\rm HI}/L_{\rm
B}~<~2\times10^{-3}$, and the evidence that it contains mostly old and
metal poor stars \citep{jerjen2000a} suggests that these passages through
the lobes may be `actively cleansing' the dwarf of any traces of accumulated
gas, probably by means of intensified ram pressure and, possibly, heating due to the higher X-ray radiation field in the plasma. In any case, this would prevent any further star formation. The other early-type dwarfs
CenA-dE1 and SGC1319.1-4216 may be in similar positions,
however, independent measurement of their distances are required to reveal
the likeliness of these objects going through the lobes.

To investigate this intensified ram pressure, we compare in Figure \ref{mtol} the $M_{\rm HI}$
and $L_{\rm B}$ values of all known galaxies from the CenA and Sculptor
group. It becomes apparent that the different global
environments, and not just the direct proximity to an AGN, acted differently
on the evolution of their respective fainter members.  When compared to the
CenA group, Sculptor seems to be lacking both low and high \hi{} mass
objects in the range $-14\,<\,M_{\rm B}\,<\,-11$.  It provides much more of
a continuum of \hi{} mass towards fainter objects than the CenA group does;
the latter seems to have more of a dichotomy between the \hi{} rich
and \hi{} deficient objects.  Precisely, Sculptor has a single galaxy with
no detected \hi{} and $M_{\rm HI}/L_{\rm B}\,<\,0.05$\,$M_{\odot}/L_{\odot,
{\rm B}}$, while, on the one hand, CenA has at least 16 galaxies that were
not detected in \hi{} and at least eight with $M_{\rm HI}/L_{\rm
B}\,<\,0.05$\,$M_{\odot}/L_{\odot, {\rm B}}$. The CenA environment
seems much more efficient in removing gas from dwarfs than the Sculptor
one.  On the other hand CenA also has a much more prominent population of
galaxies with $1\,<\,M_{\rm HI}/L_{\rm B}\,<\,10$\,$M_{\odot}/L_{\odot,{\rm
B}}$ than Sculptor.

\citet{gunn1972} determined that the ISM of a galaxy will be stripped away by
ram pressure if
\begin{equation}
\rho_{\rm IGM}~\gtrsim~\rho_{\rm ISM}\,\left(\frac{\sigma_{\rm ISM}}{v_p}\right)^2
\end{equation}
\noindent for a galaxy with an ISM of density $\rho_{\rm ISM}$ and velocity
dispersion of $\sigma_{\rm ISM}$, travelling at a velocity $v_p$ through an
intergalactic medium of density $\rho_{\rm IGM}$. By adopting $\sigma_{\rm
ISM} = 10$\,\kms{}, a value typical for the studied dwarfs
\citep[\eg][]{mateo1998, bouchard2005} and taking a value of
$v_p~\sim~300$\,\kms{}, we find that if $\rho_{\rm
IGM}~\gtrsim~10^{-3}\,\rho_{\rm ISM}$, ram pressure will eventually remove
all the gas in a low mass dwarf.  In the case of the dE ESO269-G066 the ISM
density is $\rho_{\rm ISM} < 10^{-2}\,$cm$^{-3}$ ($M_{\rm
HI}\,<\,10^5$\,\msol{} inside a presumed radius of 1\,kpc). This value is
much lower than that of the dIrr ESO324-G024, where $\rho_{\rm ISM} = 15\,$cm$^{-3}$
($M_{\rm HI}\,=\,1.5\times10^8$\,\msol{}, CFC97, inside a radius of
1\,kpc). An IGM density of the order of $\sim10^{-3}$\,cm$^{-3}$ would
therefore strip away any gas accumulation in ESO269-G066 but otherwise
leave ESO324-G024 intact.  These values for $\rho_{\rm IGM}$ are consistent
with the values measured by X-ray observations of NGC5128 \citep{cooke1978,
feigelson1981} and of loose galaxy groups \citep{mulchaey1996,
helsdon2000}. 

Figure \ref{mtol} shows that, for the faint objects of the CenA group
($M_{\rm B}\,>\,-13$), there seems to be a threshold in \hi{} masses at
values of $M_{\rm HI}\,\sim10^7$\,\msol{}.  All objects below this value
have at least a factor of 10 less \hi{} than the ones above. %Moreover, as
%there is only one detected galaxy below this limit, the dIrr ESO269-G037
%has $M_{\rm HI}\,=\,4\times10^5$\,\msol{}, the drop in \hi{} content below
%the threshold is probably more important than what we describe.  
In fact, there is only one detected galaxy below this limit: the dIrr ESO269-G037
with $M_{\rm HI}\,=\,4\times10^5$\,\msol{}.
It seems that the dwarfs situated above the threshold have kept their \hi{} while 
the ones below are efficiently swept clean of ISM. 

Assuming that the $10^7$\,\msol{} threshold indeed exists, we can redefine
the morphological classification of dwarf galaxies based on \hi{}
properties. In the $M_{\rm B}\,>\,-14$ regime of the CenA group, there are
13 late-type dwarfs ($M_{\rm HI}\,\gtrsim\,10^7$\,\msol{}) which may keep
their ISM for a long period of time.  These are objects like [CFC97]Cen6
($M_{\rm HI}/L_{\rm B}$\,=\,1.8). They have an important untapped and
stable potential for further star formation.  There is also a single
detected transition-type galaxy (ESO269-G037) that might currently be loosing its
ISM through ram pressure stripping. Finally, there are 12 early-type dwarfs
candidates, where no ISM has been detected.  Much like Ursa Minor in the
Local Group or Scl-dE1 in the Sculptor group, their lack of ISM prohibits
any further star formation and they have reached their final evolutionary
state.  If any of the latter type would contain \hi{}, it should be
considered a transition-type object because such low \hi{} content would be
short lived and prone to ram pressure stripping. The current sensitivity of \hi{}
observations allows to distinguish the late-type dwarfs but does not allow
a differentiation between the early and transition type dwarfs at the
distance of the CenA group (4.3\,Mpc).

Finally, we note that the present investigation has targeted some of the
faintest known members of the CenA group. The \hi{} survey for these
galaxies, \ie{} fainter then $M_{\rm B}\,>\,-14$, is now complete to an
\hi{} mass of $10^6$\,\msol{}. However, it is most likely that not all
dwarf galaxies of the CenA or Sculptor groups have been found to date.  For
example, in the Local Group, 10 of the $\sim40$ members are fainter than
$M_{\rm B}>-10$. In the CenA group there are only four galaxies of the
$\sim45$ galaxies that are this faint.  The advent of next generation
optical surveys will probably uncover new group members. These future
discoveries and \hi{} followup work will give an even better insight into
the properties for galaxies at the faint end of the luminosity
function, for different environments.

\section{Conclusions}

We have presented Parkes single dish and ATCA interferometric \hi{} observations of 18
low luminosity dwarf galaxies belonging to the CenA group, with
an emphasis on early-type dwarfs.
This was done in
order to identify objects with \hi{} reservoirs and, therefore, potential for
further star formation. As a result, we can constrain dwarf galaxy evolution
scenarios by providing a thorough analysis of a nearby dense environment.

The main conclusions can be summarised as follow:
\begin{enumerate}
\item{Of the 18 observed dwarfs, five were detected: three were late-type
dwarfs and two were of mixed-morphology. None of the early-type objects
were found to contain \hi{}. The detection limits on all non-detected early
and mixed-type objects are in the range from $M_{\rm HI}\,<\,10^5$\,\msol{}
to $M_{\rm HI}\,<\,6.5\times10^5$\,\msol{}, $M_{\rm HI}/L_{\rm
B}\,<\,0.24$\,$M_{\odot} L_{\odot, {\rm B}}^{-1}$.}

\item{Unlike the Local Group and the Sculptor group where all
mixed-morphology dwarfs are detected in \hi{}, only two of the four CenA
mixed-type dwarfs are detected. However, the limits on the non-detections
do not exclude \hi{} contents similar to those of other mixed-type dwarfs in
the Sculptor or Local Group.}

\item{The \hi{} masses of the detected objects range from $4\times10^5$
\msol{} to $4.2\times10^7$ \msol{}, the \hi{} mass to luminosity
ratios range from 0.04 $M_{\odot} L_{\odot, {\rm B}}^{-1}$ to 1.81
$M_{\odot} L_{\odot, {\rm B}}^{-1}$. The \hi{} velocities vary between 500
and 750 \kms{}, compatible with the expected velocities for the CenA
group members.}

\item{There is a gap in \hi{} properties for low mass CenA dwarf galaxies.
These objects either have \hi{} masses $M_{\rm HI}\,>\,10^7$\,\msol{} or
are not detected, consequently must have $M_{\rm HI}\,\lesssim\,10^6$\,\msol{}. This gap can
be explained by the ram pressure exerted by an IGM of density $\rho_{\rm
IGM}\,\sim10^{-3}$\,cm$^{-3}$.}

\item{The dE ESO269-G066 has $M_{\rm HI}/L_{\rm B}\,<\,0.002$\,$M_{\odot}
L_{\odot, {\rm B}}^{-1}$, which is much lower than the value expected from
the gas accumulated through normal stellar evolution. The orbit of the dwarf
may bring it regularly into the radio-lobes of the the active galaxy
NGC5128 constantly sweeping it free of gas.
Other galaxies may have been similarly affected by the presence of NGC5128.}

\item{An \hi{} distribution similar to what is expected from the effects of
mild ram pressure is found in UGCA365, ESO384-G016 and, possibly, in
ESO269-G037.}

\end{enumerate}

Further investigation of denser environments, such as nearby clusters, with
equally low \hi{} detection limits are required to establish whether the gap in
\hi{} properties of dwarf galaxies is common. If this is the case, it may
have profound implications on our understanding of dwarf galaxy evolution.

\begin{figure}[tbp]
\begin{center}
\includegraphics[width=0.55\textwidth]{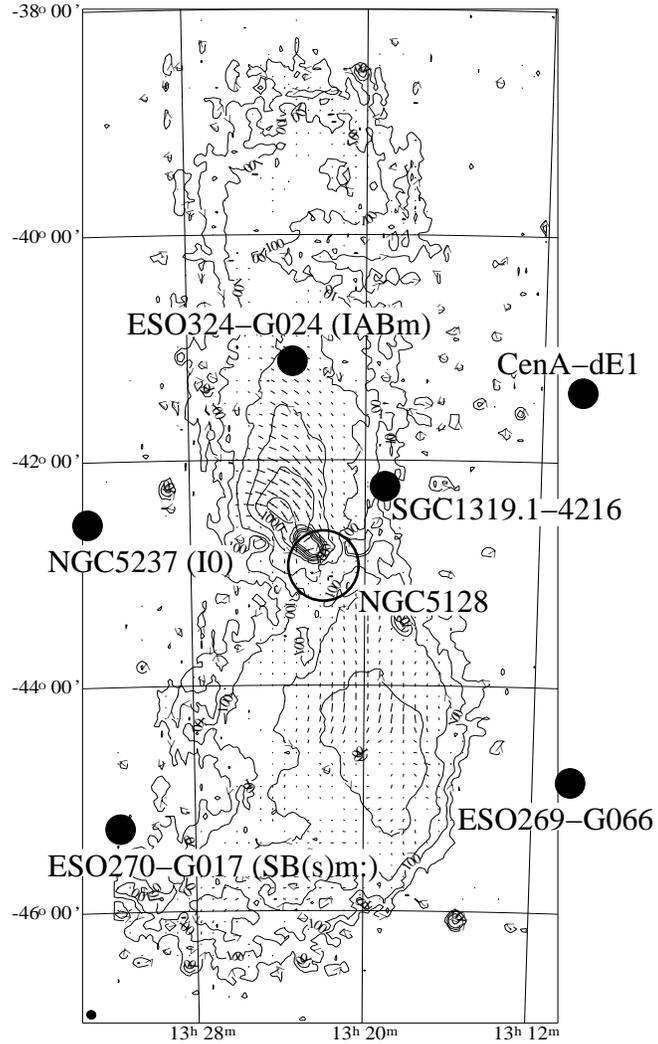}
\caption{The total power radio continuum
map around NGC5128 (\emph{open circle}), from \citet{junkes1993}. It is
overlayed with the approximate location of six neighbouring galaxies
(\emph{filled circles}). The three galaxies ESO324-G024, NGC5237 and
ESO270-G017 were not observed.  }\label{pos_cena}
\end{center}
\end{figure}

\begin{figure}[tbp]
\begin{center}
\includegraphics[width=0.95\textwidth]{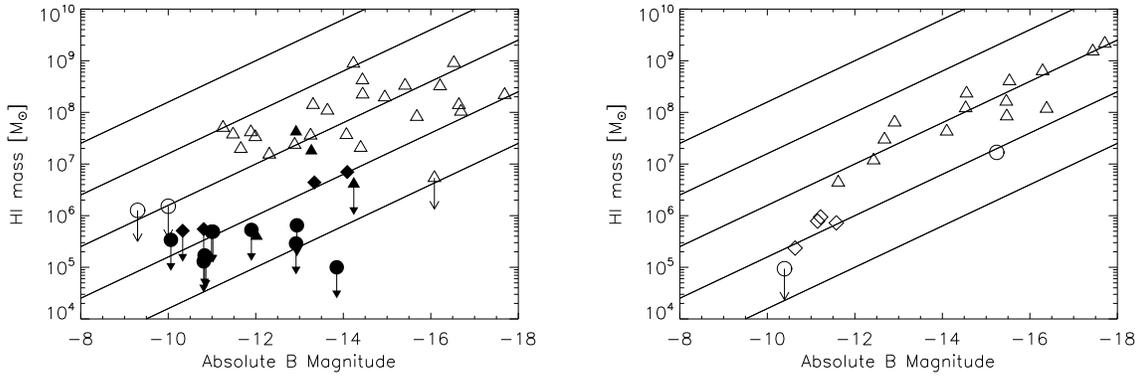}
\caption{\hi{} mass vs. absolute $B$-band magnitude for early-type dwarfs
(\emph{circles}), mixed-type dwarfs (\emph{diamonds}) and late-type dwarfs
(\emph{triangles}) of the CenA (\emph{left}) and the Scl (\emph{right})
groups. Solid symbols are the observed galaxies and the other data were
taken from \citet{cote1997, huchtmeier2000, huchtmeier2001, bouchard2005,
beaulieu2006}. The diagonal lines show constant $M_{\rm HI}/L_{\rm
B}$ values of 100, 10, 1, 0.1, 0.01 $M_{\odot}/L_{\odot, {\rm
B}}$ (\emph{top to bottom}).}\label{mtol}
\end{center}
\end{figure}

\acknowledgments
We would like to thank Sylvie Beaulieu, Geoff Bicknell and Erwin deBlok for
insightful discussions. We also thank Norbert Junkes who kindly
provided the data for Figure \ref{pos_cena} and the referee for the much appreciated comments. We acknowledge the support
from the Australian Research Council through Discovery Project Grant
DP0343156.  This research has made use of the NASA/IPAC Extragalactic
Database (NED) which is operated by the Jet Propulsion Laboratory,
California Institute of Technology, under contract with the National
Aeronautics and Space Administration. This work is partly based on
photographic data obtained using The UK Schmidt Telescope.  The UK Schmidt
Telescope was operated by the Royal Observatory Edinburgh, with funding
from the UK Science and Engineering Research Council, until 1988 June, and
thereafter by the Anglo-Australian Observatory.  Original plate material is
copyrighted by the Royal Observatory Edinburgh and the Anglo-Australian
Observatory.  The plates were processed into the present compressed digital
form with their permission.  The Digitized Sky Survey was produced at the
Space Telescope Science Institute under US Government grant NAG W-2166.

%\bibliographystyle{apj}
%\bibliography{sdw}

\clearpage

\clearpage

\clearpage

\clearpage

\clearpage

%\clearpage

\clearpage

\clearpage

\clearpage

\end{document}

%% file: tab1.tex
\begin{table*}[tbp]
\begin{footnotesize}
\begin{center}
\caption{Position and optical parameters of the sample}
\label{objects}
\begin{tabular}{l c c c c c c c c}
\hline
\hline
Galaxy				& Type	& R.A.		& Dec.		& $D$\tablenotemark{a}	& $m_{\rm B}$		& $A_{\rm B}$	& $v_{\odot}$	& References \\
				&	& (J2000)	& (J2000)	& (Mpc)			& (mag) 		& (mag)		& (km s$^{-1}$)	& 	\\
\hline
{}ESO219-G010 			& dE	& 12 56 10  	& -50 08 38	& 4.79$\pm$0.43 $s$	& 16.42$\pm$0.16	& 0.957		& \nodata	& 1, 2  \\
{}ESO269-G037			& dIrr	& 13 03 33	& -46 35 06	& 3.48$\pm$0.35 $t$	& 16.26			& 0.574		& \nodata	& 3	\\
{}[CFC97]Cen6			& dIrr	& 13 05 02	& -40 04 58	& 5.78$\pm$0.46 $t$ 	& 16.33			& 0.436		& 619$\pm$43	& 4, 5	\\
{}CenA-dE1			& dE	& 13 12 45	& -41 49 57	& 4.21$\pm$0.33 $t$ 	& 17.75$\pm$0.11	& 0.493		& \nodata 	& 1, 5	\\
{}ESO269-G066			& dE	& 13 13 09	& -44 53 24	& 4.05$\pm$0.53 $s$	& 14.59$\pm$0.08	& 0.401		& 784$\pm$31 	& 1, 2	\\
{}CenA-dE2			& dE/dIrr & 13 21 33	& -31 52 43	& \nodata 		& 18.13$\pm$0.18	& 0.288		& \nodata	& 1	\\
{}SGC1319.1-4216 		& dE	& 13 22 02	& -42 32 07	& 3.87$\pm$0.31 $t$ 	& 15.68$\pm$0.14	& 0.665		& \nodata	& 1, 5	\\
{}[CFC97]Cen8			& dIrr/dE & 13 22 56	& -33 34 22	& \nodata		& 17.65$\pm$0.08	& 0.296		& \nodata	& 1, 4	\\
{}AM1320-230			& dE	& 13 23 29	& -23 23 35	& \nodata 		& 17.53$\pm$0.08	& 0.348		& \nodata	& 1	\\
{}UGCA365			& dIrr	& 13 36 31	& -29 14 06	& 5.18$\pm$0.41 $t$ 	& 15.53			& 0.229		& 573$\pm$1	& 5, 6	\\
{}[KK98]208\tablenotemark{b}	& dIrr	& 13 36 35	& -29 34 17	& 4.68$\pm$0.42 $t$	& 14.3			& 0.192		& \nodata 	& 3	\\
{}AM1339-445 			& dE	& 13 42 05	& -45 12 18	& 3.53$\pm$0.31 $s$	& 16.32$\pm$0.1		& 0.477		& \nodata	& 2, 7  \\
{}CenA-dE3			& dE	& 13 46 00	& -36 20 15	& \nodata 		& 17.41$\pm$0.15	& 0.266		& \nodata	& 1	\\
{}AM1343-452			& dE	& 13 46 16	& -45 41 05	& 3.73$\pm$0.32 $s$	& 17.57$\pm$0.11	& 0.522		& \nodata	& 1, 2	\\
{}CenA-dE4			& dE	& 13 46 40	& -29 58 41	& \nodata 		& 17.60$\pm$0.14	& 0.260		& \nodata	& 1	\\
{}ESO384-G016			& dE/dIrr & 13 57 01	& -35 20 01	& 4.23$\pm$0.11 $s$	& 15.11$\pm$0.06	& 0.318		& 561$\pm$32	& 1, 2	\\
{}CenA-dE5			& dE	& 14 30 05	& -33 28 45	& \nodata		& 18.43$\pm$0.13	& 0.326		& \nodata	& 1	\\
{}ESO272-G025			& dE/dIrr & 14 43 25	& -44 42 18	& \nodata		& 14.77			& 0.694		& 624$\pm$10	& 3, 4	\\
\hline
\end{tabular}
\end{center}
\tablecomments{Units of right ascension are hours, minutes, and seconds, and units of declination are degrees, arcminutes, and arcseconds (J2000.0).}
\tablenotetext{a}{The distance values marked with $t$ denotes a measurement using the tip of the red
giant branch method and $s$ is used for the surface brightness fluctuation method.}
\tablenotetext{b}{[KK98]208 was not explicitly targetted but was in the field of view of UGCA365.}
\tablerefs{$A_{\rm B}$: \citet{schlegel1998}; $D$, $m_{\rm B}$: (1) \citet{jerjen2000b}; (2) \citet{jerjen2000a}; (3) \citet{karachentsev2002a}; (4) CFC97; (5) \citet{karachentsev2006b}; (6) \citet{huchtmeier2003}; (7) \citet{rejkuba2006}}
\end{footnotesize}
\end{table*}

%% file: tab2.tex
\begin{table*}[tbp]
\begin{small}
\caption{The 21\,cm observation integration time and derived physical parameters}\label{hidetections}
\begin{tabular}{l c c c c c c c}
\hline
\hline
Galaxy 		& Telescope	& T$_{\rm int}$	& $S_{\rm HI}$		& $M_{\rm HI}$		& $v_{\odot}$		& $\sigma_v$	& $M_{\rm HI}/L_{\rm B}$\\
		& 		& (min)			& (Jy km s$^{-1}$)	& ($10^5 M_{\odot}$)	& (km s$^{-1}$)	& (km s$^{-1}$)	& ($M_{\odot} L_{\odot, B}^{-1}$)	\\
\hline
{}ESO219-G010	& ATCA	& 288	& $<0.120$		& $<6.5$		& \nodata			& \nodata			& $<0.03$			\\
{}ESO269-G037	& ATCA	& 280	& 0.14$\pm$0.02	& 4$\pm$1	& 743.8$\pm$2.1	& 9.5$\pm$1.9		& 0.040$\pm$0.007	\\
{}[CFC97]Cen6		& ATCA	& 651	& 5.31$\pm$0.21	& 420$\pm$70	& 614.2$\pm$0.8	& 12.1$\pm$0.8	& 1.8$\pm$0.2		\\
{}CenA-dE1		& PKS	& 352	& $<0.038$		& $<1.6$		& \nodata			& \nodata 			& $<0.05$			\\
{}ESO269-G066	& PKS	& 651	& $<0.026$		& $<1.0$		& \nodata			& \nodata 			& $<0.002$		\\
{}CenA-dE2		& ATCA	& 271	& $<0.118$		& $<5.1$		& \nodata			& \nodata 			& $<0.24$			\\
{}SGC1319.1-4216 	& ATCA	& 614	& $<0.081$		& $<2.9$		& \nodata			& \nodata 			& $<0.01$			\\
{}[CFC97]Cen8		& ATCA	& 287	& $<0.126$		& $<5.5$		& \nodata			& \nodata 			& $<0.17$			\\
{}AM1320-230 		& ATCA	& 287	& $<0.113$		& $<4.9$		& \nodata			& \nodata 			& $<0.13$			\\
{}UGCA365		& ATCA	& 540	& 2.82$\pm$0.08	& 180$\pm$30	& 571.4$\pm$0.7	& 12.7$\pm$0.7	& 0.56$\pm$0.05	\\
{}[KK98]208		& ATCA	& 540	& $<0.80$			& $<41$		& \nodata			& \nodata 			& $<0.05$			\\
{}AM1339-445		& ATCA	& 270	& $<0.180$		& $<5.3$		& \nodata			& \nodata			& $<0.06$			\\
{}CenA-dE3		& ATCA	& 300	& $<0.112$		& $<4.9$		& \nodata			& \nodata 			& $<0.12$			\\
{}AM1343-452		& PKS	& 422	& $<0.041$		& $<1.3$		& \nodata			& \nodata 			& $<0.04$			\\
{}CenA-dE4		& PKS	& 352	& $<0.038$		& $<1.7$		& \nodata			& \nodata 			& $<0.05$			\\
{}ESO384-G016	& ATCA	& 282	& 1.04$\pm$0.18	& 44$\pm$8	& 502.9$\pm$2.4	& 8.3$\pm$2.0 		& 0.13$\pm$0.02	\\
{}CenA-dE5		& ATCA	& 529	& $<0.078$		& $<3.4$		& \nodata			& \nodata			& $<0.20$			\\
{}ESO272-G025	& ATCA	& 605	& 1.52$\pm$0.08	& 70$\pm$20	& 629.2$\pm$0.9	& 10.8$\pm$0.9	& 0.10$\pm$0.01	\\
\hline
\end{tabular}
\end{small}
\end{table*}